\documentclass[sigconf, nonacm]{acmart}
\newcommand\vldbdoi{XX.XX/XXX.XX}
\newcommand\vldbpages{XXX-XXX}
\newcommand\vldbvolume{14}
\newcommand\vldbissue{1}
\newcommand\vldbyear{2020}
\newcommand\vldbauthors{\authors}
\newcommand\vldbtitle{\shorttitle} 
\newcommand\vldbavailabilityurl{URL_TO_YOUR_ARTIFACTS}
\newcommand\vldbpagestyle{plain} 

\newcommand{\eat}[1]{}

\usepackage{latexsym}
\usepackage{amsthm}
\usepackage{xcolor}
\usepackage{colortbl}
\usepackage{epsfig}
\usepackage{xspace}
\usepackage{graphicx}
\usepackage{subfigure}
\usepackage{paralist}
\usepackage{enumitem}
\usepackage[color,matrix,arrow,all]{xy}
\usepackage{comment}
\usepackage{booktabs}
\usepackage{balance}
\usepackage{stmaryrd}
\usepackage{pifont}
\usepackage{hhline}
\usepackage{listings}
\usepackage{array}
\usepackage{float}
\usepackage[flushleft]{threeparttable}

\usepackage{mathrsfs}
\usepackage{makecell}
\usepackage{xparse}
\usepackage{wrapfig}

\usepackage{epsfig}
\usepackage{multirow}
\usepackage{url}

\usepackage{multirow}
\usepackage{natbib}
\usepackage{graphicx}

\usepackage{listings}
\usepackage{framed}
\usepackage{xcolor}
\usepackage{color}
\usepackage{geometry}
\usepackage{changepage}
\colorlet{shadecolor}{gray!20}

\definecolor{shadecolor}{RGB}{220,220,220}

\definecolor{inputcolor}{RGB}{255,139,35}
\definecolor{outputcolor}{RGB}{120,212,252}
\definecolor{embedcolor}{RGB}{254,127,156}
\definecolor{maskcolor}{RGB}{122,128,255}
\definecolor{ecolor}{RGB}{58,149,54}

\definecolor{highcolor}{RGB}{255,153,153}
\definecolor{midcolor}{RGB}{255,204,204}
\definecolor{lowcolor}{RGB}{204,229,255}

\newtheorem{example}{Example}

\usepackage{tikz}
\usetikzlibrary{shapes,snakes}
\usetikzlibrary{calc}

\usepackage[export]{adjustbox}

\definecolor{green}{RGB}{0,128,0}

\definecolor{yellow}{RGB}{255,200,18}

\newcommand{\bi}{\begin{itemize}}
\newcommand{\ei}{\end{itemize}}

\newcommand{\be}{\begin{enumerate}}
\newcommand{\ee}{\end{enumerate}}
\newcommand{\beqn}{\begin{eqnarray*}}
\newcommand{\eeqn}{\end{eqnarray*}}

\newcommand{\stitle}[1]{\vspace{1mm}\noindent{\bf #1}}
\newcommand{\etitle}[1]{\vspace{0.5mm}\noindent{\underline{\em #1}}}

\newcommand{\sys}{\texttt{RaG-Tree}\xspace}

\usepackage{amsmath, amsthm}
 
\NewDocumentCommand{\nan}{ mO{} }{\textcolor{red}{\textsuperscript{\textit{Nan}}\textsf{\textbf{\small[#1]}}}}

\usepackage{algorithm}
\usepackage{algorithmicx}
\usepackage{algpseudocode}
\usepackage{amsmath} 
\floatname{algorithm}{Algorithm} 

\NewDocumentCommand{\cc}{mO{}}{\textcolor{blue}
{\textsuperscript{\textit{CC}}\textsf{\textbf{\small[#1]}}}}

\NewDocumentCommand{\fanj}{mO{}}{\textcolor{orange}
	{\textsuperscript{\textit{fanj}}\textsf{\textbf{\small[#1]}}}}

\usepackage{cases}
\usepackage{cleveref}
\usepackage{makecell}
\usepackage{bm}

\usepackage{tablefootnote}

\usepackage{arydshln}

\definecolor{darkred}{rgb}{0.75, 0.0, 0.0}
\definecolor{darkgreen}{rgb}{0.0, 0.45, 0.0}

\definecolor{mint}{rgb}{0.62, 0.89, 0.75}
\definecolor{lightrose}{rgb}{0.93, 0.26, 0.22}
\definecolor{blu}{rgb}{0.18359375, 0.4296875, 0.7265625}

\colorlet{mint5}{mint!5}
\colorlet{mint10}{mint!10}
\colorlet{mint20}{mint!20}
\colorlet{mint30}{mint!30}
\colorlet{mint40}{mint!40}
\colorlet{mint50}{mint!50}
\colorlet{mint60}{mint!60}
\colorlet{mint70}{mint!70}
\colorlet{mint80}{mint!80}
\colorlet{mint90}{mint!90}
\colorlet{mint100}{mint!100}

\colorlet{rose5}{lightrose!5}
\colorlet{rose10}{lightrose!10}
\colorlet{rose20}{lightrose!20}
\colorlet{rose21}{lightrose!21}
\colorlet{rose30}{lightrose!30}
\colorlet{rose40}{lightrose!40}
\colorlet{rose45}{lightrose!45}
\colorlet{rose50}{lightrose!50}
\colorlet{rose60}{lightrose!60}
\colorlet{rose70}{lightrose!70}
\colorlet{rose80}{lightrose!80}
\colorlet{rose90}{lightrose!90}
\colorlet{rose100}{lightrose!100}

\colorlet{green5}{green!5}
\colorlet{green10}{green!10}
\colorlet{green20}{green!20}
\colorlet{green30}{green!30}
\colorlet{green40}{green!40}
\colorlet{green50}{green!50}
\colorlet{green60}{green!60}
\colorlet{green70}{green!70}
\colorlet{green80}{green!80}
\colorlet{green90}{green!90}
\colorlet{green100}{green!100}

\colorlet{blu5}{blu!5}
\colorlet{blu10}{blu!10}
\colorlet{blu20}{blu!20}
\colorlet{blu30}{blu!30}
\colorlet{blu40}{blu!40}
\colorlet{blu50}{blu!50}
\colorlet{blu60}{blu!60}
\colorlet{blu70}{blu!70}
\colorlet{blu80}{blu!80}
\colorlet{blu90}{blu!90}
\colorlet{blu100}{blu!100}

\colorlet{orange5}{orange!5}
\colorlet{orange10}{orange!10}
\colorlet{orange20}{orange!20}
\colorlet{orange30}{orange!30}
\colorlet{orange40}{orange!40}
\colorlet{orange50}{orange!50}
\colorlet{orange60}{orange!60}
\colorlet{orange70}{orange!70}
\colorlet{orange80}{orange!80}
\colorlet{orange90}{orange!90}
\colorlet{orange100}{orange!100}

\usepackage{scalerel}
\usepackage{stackengine}
\usepackage{pgf}
\newcounter{iloop}
\newcommand\openbigstar[1][0.7]{%
  \scalerel*{%
    \stackinset{c}{-.125pt}{c}{}{\scalebox{#1}{\color{white}{$\bigstar$}}}{%
      $\bigstar$}%
  }{\bigstar}
}
\newcommand{\Stars}[1]{\ensuremath{%
\pgfmathtruncatemacro{\imax}{ifthenelse(int(#1)==#1,#1-1,#1)}%
\pgfmathsetmacro{\xrest}{0.9*(1-#1+\imax)}%
\setcounter{iloop}{0}%
\loop\stepcounter{iloop}\ifnum\value{iloop}<\the\numexpr\imax+1
\bigstar\repeat
\openbigstar[\xrest]%
\setcounter{iloop}{0}%
\loop\stepcounter{iloop}\ifnum\value{iloop}<\the\numexpr5-\imax\relax
\openbigstar[.9]\repeat}}

\newcommand{\ovec}{\mathbf{x}\xspace}
\newcommand{\oattr}{\mathbf{a}\xspace}
\newcommand{\qvec}{\mathbf{q}\xspace}
\newcommand{\qattr}{\mathcal{R}\xspace}
\newcommand{\vhnsw}{{G}\xspace}
\newcommand{\hnsw}{HNSW\xspace}
\newcommand{\mbr}{MBR\xspace}

\begin{document}

\title{RaG-Tree: Combining R-Tree and HNSW for Multi-Attribute Range Filtered Approximate Nearest Neighbor Search}

\settopmatter{authorsperrow=3}
\author{Jiawei Liu}
\affiliation{%
	\institution{Renmin University of China}
}
\email{jiaweiliu@ruc.edu.cn}

\author{Xiang Zhang}
\affiliation{%
	\institution{Renmin University of China}
}
\email{xiangzhang@ruc.edu.cn}

\author{Chao Zhang}
\affiliation{%
	\institution{Renmin University of China}
}
\email{cycchao@ruc.edu.cn}

\author{Ju Fan}
\affiliation{%
	\institution{Renmin University of China}
}
\email{fanj@ruc.edu.cn}


\author{Xiaoyong Du}
\affiliation{%
	\institution{Renmin University of China}
}
\email{duyong@ruc.edu.cn}

\begin{abstract}
%
Multi-attribute range-filtered approximate nearest neighbor search (MR-ANNS), which retrieves high-dimensional vectors satisfying multiple attribute constraints, is a fundamental operation in modern AI applications. Existing MR-ANNS indexes either exploit a single attribute for range localization or recursively partition objects along individual attributes, which may limit their ability to exploit attribute correlations for effective range pruning and attribute-vector correlations for efficient nearest-neighbor search.
In this paper, we propose \sys, a unified index that couples an R-tree with partition-aware HNSW graphs for MR-ANNS. \sys leverages hierarchical R-tree partitions for effective range pruning and adapts the sparsity of each HNSW graph to the local vector distributions within its partition, enabling lightweight indexing and efficient query processing. To support efficient query processing and dynamic updates, we develop a cost-based adaptive search algorithm that minimizes unnecessary graph exploration, together with an efficient index maintenance mechanism for incrementally updating affected partition-aware HNSW graphs.
Extensive experiments on three real-world datasets show that \sys achieves superior query performance over state-of-the-art baselines, while also providing lightweight indexing and fast incremental updates.

\end{abstract}

\maketitle

\pagestyle{\vldbpagestyle}
\begingroup\small\noindent\raggedright\textbf{PVLDB Reference Format:}\\
\vldbauthors. \vldbtitle. PVLDB, \vldbvolume(\vldbissue): \vldbpages, \vldbyear.\\
\href{https://doi.org/\vldbdoi}{doi:\vldbdoi}
\endgroup
\begingroup
\renewcommand\thefootnote{}\footnote{\noindent
	This work is licensed under the Creative Commons BY-NC-ND 4.0 International License. Visit \url{https://creativecommons.org/licenses/by-nc-nd/4.0/} to view a copy of this license. For any use beyond those covered by this license, obtain permission by emailing \href{mailto:info@vldb.org}{info@vldb.org}. Copyright is held by the owner/author(s). Publication rights licensed to the VLDB Endowment. \\
	\raggedright Proceedings of the VLDB Endowment, Vol. \vldbvolume, No. \vldbissue\ %
	ISSN 2150-8097. \\
	\href{https://doi.org/\vldbdoi}{doi:\vldbdoi} \\
}\addtocounter{footnote}{-1}\endgroup

\ifdefempty{\vldbavailabilityurl}{}{
	\vspace{.3cm}
	\begingroup\small\noindent\raggedright\textbf{PVLDB Artifact Availability:}\\
	The source code, data, and/or other artifacts have been made available at \url{https://github.com/rucjrliu/RaG-Tree_code}.
	\endgroup
}

\section{Introduction}
\label{sec:introduction}

Nearest neighbor search (NNS), which retrieves the objects closest to a query vector, has become a fundamental operation in modern AI systems, including recommendation, information retrieval, and retrieval-augmented generation~\cite{chen2022approximate,lewis2020retrieval,khattab2020colbert,jiang2025piperag}. As exact NNS is prohibitively expensive over large-scale and high-dimensional vector data, approximate nearest neighbor search (ANNS) is proposed by trading accuracy for search efficiency, and has been widely adopted by modern vector search systems~\cite{zhang2026vecbench,huang2026survey,PGVector,wang2021milvus}.

In real-world applications, each object is typically associated with not only \emph{a high-dimensional vector} but also \emph{multiple scalar attributes}. Queries therefore need to retrieve the top-$k$ nearest neighbors while satisfying user-specified range predicates over these attributes. We refer to this problem as \emph{Multi-attribute Range-filtered Approximate Nearest Neighbor Search} (MR-ANNS). For example, an e-commerce query~\cite{amazon,taobao} may retrieve products semantically similar to a given image while constraining price, popularity, and inventory to specified ranges. Efficient support for MR-ANNS is thus essential for practical vector retrieval over real-world structured datasets.

\stitle{Limitations of Existing Methods.}
Most existing range-filtered ANNS indexes are organized around a \emph{single filtering attribute}~\cite{jiang2025digra,xu2024irangegraph,zuo2024serf,zhang2025rangepq,liang2024unify}. Given a query range, the methods first localize the search to the corresponding attribute interval and then perform ANNS within the resulting search space. However, under multi-attribute range queries, only one attribute can be exploited to localize the search, while the remaining attributes are evaluated during or after ANNS. Consequently, the search explores a large number of \emph{out-of-range} objects that lie outside the multi-attribute query region, significantly affecting query efficiency.

Recently, KHI~\cite{yu2026efficient} extends range-filtered ANNS to multi-attribute queries by recursively partitioning objects along individual attributes using a KD-tree and constructing an HNSW graph~\cite{malkov2014approximate} for each resulting object partition. Given an MR-ANNS query, it recursively traverses the KD-tree to identify the relevant partitions and performs ANNS on the corresponding HNSW graphs. However, this design has the following two limitations.
First, recursively partitioning objects along individual attributes overlooks \emph{attribute correlations}, leading to inefficient range localization and causing a query to search unnecessarily many HNSW graphs.
Second, constructing HNSW graphs without considering various \emph{attribute-vector correlations} across partitions leads to unnecessary graph exploration over out-of-range objects.

\stitle{Our Proposal.}
To address these limitations, we propose \sys, a unified index that tightly couples an R-tree with partition-aware HNSW graphs for efficient MR-ANNS. First, \sys organizes objects using an R-tree built over all attributes, preserving attribute correlations through hierarchical multi-attribute partitions. This enables accurate range localization via Minimum Bounding Rectangle (MBR) pruning, thereby pruning irrelevant object partitions early and significantly reducing the number of HNSW graphs searched for each query. Second, \sys constructs a partition-aware HNSW graph for each R-tree node, where graph sparsity is adaptively determined according to attribute-vector correlations, reducing unnecessary graph exploration.



\stitle{Challenges and Solutions.}
Realizing \sys requires addressing several challenges in query processing and index maintenance. 

\etitle{Query Processing.}
Given an MR-ANNS query, \sys first identifies the R-tree nodes intersecting with the query predicates and then executes ANNS on the corresponding HNSW graphs. The key challenge is determining the appropriate combination of intersected R-tree nodes across different levels to minimize the overall search cost. To illustrate this trade-off, we consider two baseline strategies. Searching only higher-level nodes executes a few HNSW graphs but incurs excessive search over out-of-range objects. Searching only lower-level nodes improves object filtering but requires executing many HNSW graphs. To address this, we propose a \emph{cost-based adaptive search} method that adaptively selects the optimal combination of intersected R-tree nodes across different levels, minimizing the overall search cost and substantially reducing query latency.


\etitle{Index Updates.}
Index updates maintain the R-tree and the partition-aware HNSW graphs under object insertions and deletions. While the R-tree can be updated efficiently through incremental MBR maintenance, efficiently maintaining the HNSW graphs remains challenging. Existing MR-ANNS indexes such as KHI~\cite{yu2026efficient} do not support dynamic updates. The key challenge is efficiently propagating each update to multiple HNSW graphs, as an object may belong to multiple R-tree nodes.
To address this challenge, we propose a delta HNSW graph mechanism. Insertions are buffered in compact delta HNSW graphs, and deletions are efficiently tracked using bitmaps. Delta HNSW graphs are then adaptively merged into the main HNSW graphs for frequently searched nodes, enabling efficient updates while maintaining high query performance.


\stitle{Contributions.} Our contributions are summarized as follows. 
%
\begin{itemize}[leftmargin=*]
\item We propose \sys, a unified index that couples an R-tree with partition-aware HNSW graphs for MR-ANNS, leveraging R-tree partitions for efficient range pruning and tailoring graph sparsity to attribute-vector correlations across partitions, yielding lightweight indexing and efficient query processing (Section~\ref{sec:overview}).



\item We propose novel algorithms for query processing (Section~\ref{sec:online-search}) and index maintenance (Section~\ref{sec:index-update}) tailored to \sys, enabling low-latency query execution and efficient dynamic updates.

%


\item We conduct extensive experiments on well-adopted MR-ANNS benchmarks (Section~\ref{sec:experiments}). The results show that, at around $0.95$ recall, \sys improves query throughput by up to $2.8\times$ over baselines, with the advantage growing to more than $10\times$ as the number of attributes increases, while maintaining efficient index construction, lightweight index size and fast updates.


\end{itemize}

\section{Preliminaries} \label{sec:problem}


%




This section first defines the MR-ANNS problem (Section~\ref{subsec:problem}) and then briefly introduces R-tree and HNSW (Sections~\ref{subsec:r-tree} and \ref{subsec:hnsw}).

\subsection{Problem Formulation} \label{subsec:problem}


\stitle{Data.}
This paper considers a set of objects $O=\{o_i \mid 1 \le i \le n\}$ associated with $m$ attributes $\{A_1,\ldots,A_m\}$.
Specifically, each object $o_i=(\ovec_i,\oattr_i)$ consists of a $d$-dimensional vector $\ovec_i$ and its attribute values $\oattr_i=(a_{i1},\ldots,a_{im})$, where $a_{ij}$ denotes the value of object $o_i$ on attribute $A_j$.
Without loss of generality, we assume all attributes are numeric, since categorical and datetime attributes can be encoded as numeric values.

\begin{figure*}[t]
	\centering
	\hspace{-1em}
	\subfigure[\sys Structure and Query Processing.]{
		\includegraphics[width=0.68\textwidth]{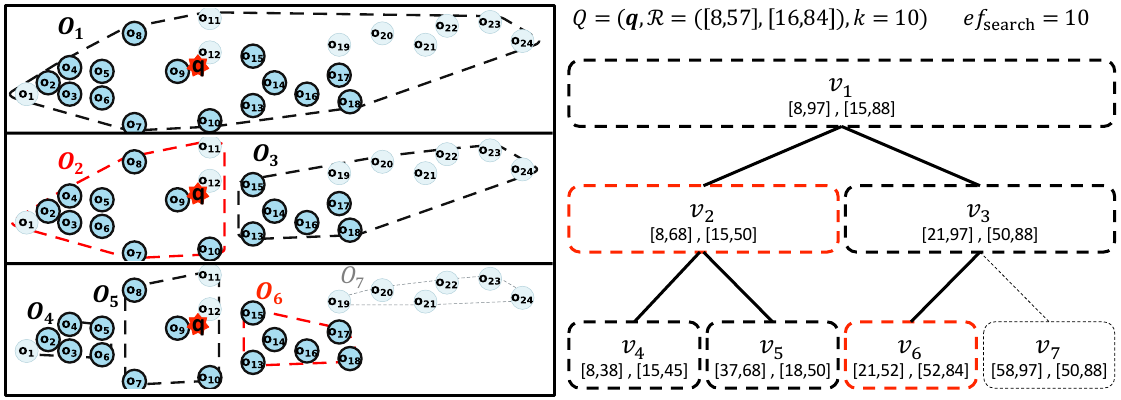}
		\label{fig:rtreehnsw-1}
	} 
	\subfigure[Local \hnsw Graphs.]{
		\includegraphics[width=0.29\textwidth]{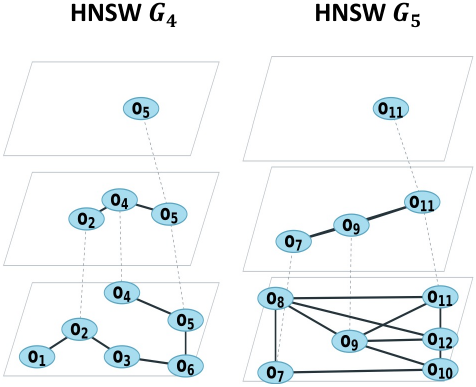}
		\label{fig:rtreehnsw-2}
	} 
    \vspace{-1em}
    \caption{An overview of \sys, including its R-tree-based index structure and query processing. Each node \(v\) corresponds to an object subset \(O_v\) and maintains a local HNSW graph \(\vhnsw_v\) constructed over \(O_v\).}
	\label{fig:rtreehnsw}
\end{figure*}

\stitle{Query.}
An MR-ANNS query is represented as $Q=(\qvec,\qattr,k)$, where $\qvec$ is a $d$-dimensional query vector, $\qattr=(\qattr_1,\ldots,\qattr_m)$ is a set of query ranges over attributes $\{A_1,\ldots,A_m\}$, and $k$ is the number of nearest neighbors to retrieve.
Specifically, each query range $\qattr_j=[l_j,u_j]$ specifies the range constraint on attribute $A_j$.

\stitle{The MR-ANNS Problem.}
The objects satisfying all query range constraints form the candidate object set
\[
O(\qattr)=
\{o_i\in O \mid a_{ij}\in \qattr_j,\ \forall\,1\le j\le m\}.
\]
Moreover, we use the Euclidean distance $\|\ovec_i-\qvec\|$ to measure the similarity between an object vector $\ovec_i$ and the query vector $\qvec$.

Based on the definitions above, we are ready to define the MR-ANNS problem as follows.
\begin{definition}[MR-ANNS]
Given an object set $O$ and a query $Q=(\qvec,\qattr,k)$, let $S_k(Q)$ denote the exact top-$k$ nearest neighbors of $\qvec$ among all objects in $O(\qattr)$. The MR-ANNS problem aims to return an approximate top-$k$ result $\widehat{S}_k(Q)$.
\end{definition}

For example, consider the MR-ANNS query $Q=(\qvec,\qattr,k)$ illustrated in \Cref{fig:rtreehnsw-1}, where $\qvec$ is the query vector, $\qattr=([8,57],[16,84])$, and $k=10$.
The attribute constraints first define a candidate object set consisting of all objects satisfying $\oattr_1\in[8,57]$ and $\oattr_2\in[16,84]$. 
Among these candidate objects, the MR-ANNS query returns the approximate top-$10$ nearest neighbors to $\qvec$.

\subsection{R-Tree} \label{subsec:r-tree}

An R-tree is a height-balanced spatial index that recursively partitions objects in the multi-dimensional attribute space~\cite{guttman1984rtree}. Each node summarizes the objects in its subtree using a Minimum Bounding Rectangle (MBR), enabling efficient pruning of irrelevant partitions during range query processing.

Formally, let $T_{v_1,V}$ denote an R-tree rooted at node $v_1$, where $V$ is the set of tree nodes. For each node $v\in V$, let $O_v\subseteq O$ denote the objects contained in the subtree rooted at $v$. The MBR of node $v$, denoted by $R_v=(R_{v,1},\ldots,R_{v,m})$, bounds the values of all $m$ attributes over $O_v$, where
\[
R_{v,j}=[l_j^{R_v},u_j^{R_v}]
\]
with
\[
l_j^{R_v}=\min\{a_{ij}\mid o_i\in O_v\},\qquad
u_j^{R_v}=\max\{a_{ij}\mid o_i\in O_v\}.
\]
Consequently, every object $o_i\in O_v$ satisfies $a_{ij}\in R_{v,j}$ for all $1\le j\le m$.
Given a query $Q=(\qvec,\qattr,k)$, the R-tree traverses only the nodes whose MBRs intersect the query ranges, i.e.,
\[
R_{v,j}\cap \qattr_j\neq\emptyset,\quad\forall\,1\le j\le m.
\]
After MBR pruning, let \(O_v(\qattr)\) denote the subset of objects in \(O_v\) satisfying the query constraints.

\subsection{\hnsw} \label{subsec:hnsw}

\hnsw is a hierarchical graph-based index for approximate nearest neighbor search (ANNS)~\cite{malkov2018efficient}. It organizes an object set $O$ into a hierarchy of proximity graphs, where each vertex represents an object and each edge connects two neighboring objects according to their vector similarity. The upper layers contain progressively fewer vertices for coarse-grained navigation, while the bottom layer contains all objects for fine-grained nearest-neighbor search.

Formally, let $\vhnsw$ denote the HNSW index built over object set $O$. The graph sparsity is controlled by the maximum degree parameter $M$, which specifies the maximum number of neighbors maintained for each vertex. During index construction, each vertex is assigned a random maximum layer, yielding a hierarchical graph whose expected height is approximately $h=\lfloor\log_{M}|O|\rfloor+1$.
Given a query vector $\qvec$, HNSW greedily descends from the top layer to the
bottom layer and then performs a best-first graph search on the bottom layer.
The search is controlled by the parameter $ef_{\mathrm{search}}$, which specifies
the maximum size of the candidate list maintained during graph exploration,
trading off search accuracy and efficiency.
\section{\sys} \label{sec:overview}

This section first presents the overall design of the proposed \sys index (Section~\ref{subsec:index-overview}), followed by an efficient algorithm for constructing the index (Section~\ref{subsec:index-construct}).

\subsection{Index Overview} \label{subsec:index-overview}

We propose \sys, a unified index that tightly couples an R-tree with partition-aware HNSW graphs for efficient MR-ANNS.

\stitle{Design Principles.}
The design of \sys is motivated by two complementary properties of multi-attribute vector data.

First, objects often exhibit strong \emph{attribute correlations}, where multiple attributes jointly determine the data distribution in the attribute space.
For example, products with similar prices are often associated with similar brands or categories, while houses in nearby locations tend to have similar sizes and prices.
Due to such correlations, an R-tree can partition attribute space into compact regions with tight MBRs, thereby improving range pruning efficiency.

Second, objects exhibit \emph{attribute-vector correlations}, meaning that the proximity among vectors varies across different regions of the attribute space.
For example, vectors associated with one attribute region may be tightly clustered with small pairwise distances, whereas those associated with another region may be substantially more dispersed.
Due to such correlations, the graph sparsity should be adapted to the local vector proximity of each attribute partition, thereby constructing more compact graphs while maintaining efficient ANNS.

Therefore, the above two properties motivate a unified index design: attribute correlations are exploited to organize the hierarchical partition structure, while attribute-vector correlations are leveraged to optimize the vector index within each partition.

\stitle{Index Structure.}
As illustrated in \Cref{fig:rtreehnsw-1}, \sys is a hierarchical MR-ANNS index that couples an R-tree with partition-aware \hnsw graphs. The R-tree recursively partitions the object set according to attribute values, while each tree node serves as a unified indexing unit for both attribute filtering and vector search by maintaining a local \hnsw graph over the objects covered by the node.

Formally, \sys constructs a binary R-tree $T_{v_1,V}$ with node set $V$, rooted at node $v_1\in V$.
Each node $v\in V$ is represented by a 5-tuple $(O_v,R_v,\vhnsw_{v},lch_v,rch_v)$,
where $O_v$ denotes the object subset covered by $v$, $R_v$ is the minimum bounding rectangle (MBR) of $O_v$ over all $m$ attributes, $\vhnsw_{v}$ is the local \hnsw graph built over $O_v$, and $lch_v$ and $rch_v$ denote the left and right child nodes, respectively. In examples involving a specific node \(v_i\), we simplify the notation by writing \((O_i,R_i,\vhnsw_{i},lch_i,rch_i)\).

Spcifically, for each node $v$, \sys constructs a partition-aware \hnsw graph $\vhnsw_{v}$ over the corresponding object set $O_v$. For ease of presentation, we use \emph{node} to refer to an R-tree node and \emph{vertex} to refer to an HNSW vertex. Thus, each each vertex in $\vhnsw_{v}$ corresponds to an object $o\in O_v$, while each edge connects neighboring objects according to their Euclidean distances in the vector space. 
In particular, each local \hnsw graph $\vhnsw_{v}$ is associated with a maximum degree parameter $M_v$,
which controls its graph sparsity. Since each tree node corresponds to a distinct attribute partition,
different local graphs may adopt different values of $M_v$ to better capture the attribute-vector correlations across partitions. 

The above hierarchical organization enables \sys to jointly exploit attribute correlations for effective range pruning and attribute-vector correlations for localized vector indexing.

\begin{example}
\label{eg:overviewlocalgraph}
To illustrate the index structure, consider the example in \Cref{fig:rtreehnsw-1}. The object set $O$ is recursively partitioned in the attribute space, e.g., $O=O_1=O_2\cup O_3$ and $O_2=O_4\cup O_5$. Each tree node maintains an MBR that bounds the attribute region of its object subset, such as $R_2=([8,68],[15,50])$.
As illustrated in \Cref{fig:rtreehnsw-2}, each R-tree node maintains a local \hnsw graph for MR-ANNS. Each local graph is associated with a maximum degree parameter $M_v$, which determines its graph sparsity. Since different attribute partitions exhibit different local vector distributions, different graphs may adopt different values of $M_v$. For example, the local graphs associated with $O_4$ and $O_5$ use different maximum degrees ($M_4=2$ and $M_5=4$), resulting in different graph sparsities. 
\end{example}
To balance index size, construction cost, and ANNS performance, \sys adaptively determines the maximum degree $M_v$ of each local graph through our proposed index construction method presented in Section~\ref{subsec:index-construct}.


\stitle{Query Processing.}
%
%
%
%
Given a query $Q=(\qvec,\qattr,k)$, \sys first performs \mbr{} pruning on the R-tree to identify all predicate-intersecting nodes. It then selects a subset of these nodes whose \mbr{}s collectively cover the query predicates and performs coordinated MR-ANNS over their local \hnsw{} graphs.

Specifically, for each selected node $v$, \sys performs MR-ANNS on the local \hnsw{} graph $\vhnsw_v$ with a local search parameter $ef_v$. The search starts from the entry point at the top layer of $\vhnsw_v$ and greedily traverses each layer toward vertices closer to $\qvec$, where the closest vertex found at the current layer serves as the entry point for the next lower layer. Upon reaching the bottom layer, \sys maintains a candidate list of size $ef_v$ and iteratively expands the most promising vertices according to their distances to $\qvec$. Therefore, $ef_v$ controls the search breadth of the local graph: a larger value explores more vertices, typically improving recall at the cost of higher search overhead. During the traversal, all visited vertices can be used for graph navigation, whereas only objects satisfying the query predicates $\qattr$ are retained as result candidates.

Finally, \sys merges candidates returned from the selected local \hnsw{} graphs and obtains the approximate top-$k$ result $\widehat{S}_k(Q)$.

\begin{example}
\label{eg:overviewqp}
%
In \Cref{fig:rtreehnsw-1}, given the query $Q=(\qvec,\qattr=([8,57],[16,84]),k=10)$, \sys first performs \mbr{} pruning and identifies the range-intersecting nodes $V(\qattr)=\{v_1,v_2,v_3,v_4,v_5,v_6\}$, while node $v_7$ is pruned.
The selected nodes for local graph search should ensure that the union of their \mbr{}s covers the query predicates $\qattr$, yielding multiple feasible node combinations, such as $\{v_1\}$, $\{v_2,v_6\}$, and $\{v_4,v_5,v_6\}$.

Different node combinations exhibit different search behaviors. For example, selecting only $v_1$ performs MR-ANNS on a single large local graph, where many visited objects do not satisfy the query ranges. In contrast, selecting all intersecting leaf nodes, i.e., $\{v_4,v_5,v_6\}$, reduces unnecessary filtering but requires searching three local graphs. As illustrated in \Cref{fig:rtreehnsw-1}, \sys eventually performs coordinated MR-ANNS on the local graphs associated with $v_2$ and $v_6$, enabling an effective balance between the number of searched local graphs and the number of visited irrelevant objects.
\end{example}

%

The above example illustrates two key challenges in query processing. 
First, multiple node combinations may satisfy the query ranges, but exhibit different trade-offs between searching more local graphs and visiting more irrelevant objects. 
Second, the user-provided search parameter $ef_{\mathrm{search}}$ should be appropriately allocated among the selected local graphs. 
To address the challenges, we propose a \emph{cost-based adaptive search} method that adaptively selects the optimal combination of intersected R-tree nodes across different levels, minimizing the overall search cost and substantially reducing query latency. Please refer to Section~\ref{sec:online-search} for more details.

\subsection{\hnsw Construction} \label{subsec:index-construct}

\begin{algorithm}[t]
\caption{\sys{} Construction}
\label{alg:ragtreeconstruction}
\begin{algorithmic}[1]
\Require Object set $O$, Adaptive maximum-degree range $[M_L,M_H]$
\Ensure Constructed \sys $T_{v_1,V}$
\Function{ConstructNode}{$O_v$}
    \State $(v,R_v,O_{lch_v},O_{rch_v})
           \gets\Call{PartitionAndCalc\mbr}{O_v}$
    \State $M_v\gets
           \Call{CalcMaxDeg}{O_v,R_v,[M_L,M_H]}$
    \State $\vhnsw_v\gets
           \Call{Construct\hnsw{}Graph}{O_v,M_v}$
    \If{$O_{lch_v}\neq\emptyset\land O_{rch_v}\neq\emptyset$} 
        \State $lch_v\gets
               \Call{ConstructNode}{O_{lch_v}}$ 
        \State $rch_v\gets
               \Call{ConstructNode}{O_{rch_v}}$
    \EndIf
    \State \Return $v$
\EndFunction
\State $v_1\gets\Call{ConstructNode}{O}$
\State $V\gets\{v_1,\ldots,v_{|V|}\}$ by Traverse from Root $v_1$
\State \Return $T_{v_1,V}$
\end{algorithmic}
\end{algorithm}


We recursively construct an \sys index over the object set $O$ from the root node, as outlined in~\Cref{alg:ragtreeconstruction}. For each node $v$, \sys performs the following three steps.
\begin{itemize}[leftmargin=*]
\item \textbf{Step 1:} Partitioning the object subset $O_v$ and computing the corresponding \mbr{} using \emph{PartitionAndCalcMBR}.
\item \textbf{Step 2:} Adaptively determining the maximum degree $M_v$ of the local \hnsw{} graph according to the vector distributions within the attribute partition represented by node $v$.
\item \textbf{Step 3:} Constructing the local \hnsw{} graph $\vhnsw_v$ using \emph{Construct\hnsw{}Graph}, and recursively repeating the same procedure for the child nodes.
\end{itemize}

The main challenge lies in the second step, determining an appropriate maximum degree $M_v$ for each local \hnsw{} graph $\vhnsw_v$, as different attribute partitions exhibit different local vector distributions.
The key observation is that a local partition with smaller pairwise vector distances requires fewer graph connections, whereas one with larger pairwise distances requires a denser local graph. Therefore, \sys determines the maximum degree of each local graph according to its local vector distribution.

\stitle{Measuring Local Vector Distributions.}
To characterize the local vector distribution, \sys uses the average pairwise vector distance (APD) among the objects in a partition. Specifically, the APD of an object set $O$ is defined as
\begin{equation}
\mathrm{APD}(O)
=
\frac{1}{\binom{|O|}{2}}
\sum_{\{o_i,o_j\}\subset O}
\|\ovec_i-\ovec_j\|.
\end{equation}

For a node \(v\) associated with the object subset \(O_v\), \sys 
normalizes its APD by that of the entire object set $O$, i.e., 
\begin{equation}
\label{eq:def3}
\delta_O(O_v)=
\frac{\mathrm{APD}(O)-\mathrm{APD}(O_v)}
{\mathrm{APD}(O)}
\in(-\infty,1].
\end{equation}

A larger value of $\delta_O(O_v)$ indicates that vectors in $O_v$ have smaller pairwise distances than those in the entire set $O$, whereas a smaller or negative value indicates larger pairwise distances.

\stitle{Adaptive Maximum Degree Determination.}
Based on the relative change in APD, \sys linearly maps the non-negative value of $\delta_O(O_v)$ to the predefined maximum-degree range $[M_L,M_H]$:
\begin{equation}
\label{eq:adahnswmaxdeg}
M_v=
\mathrm{Round}
\left(
M_H-
\max\{0,\delta_O(O_v)\}
(M_H-M_L)
\right).
\end{equation}

Based on $\delta_O(O_v)$, each local partition is assigned an adaptive maximum degree within the predefined range $[M_L,M_H]$. The resulting maximum degree is then used to construct the corresponding local \hnsw{} graph, whose height follows the default HNSW setting:
\[
h_v=\left\lceil\log_{M_v}|O_v|\right\rceil.
\]

Note that computing $\delta_O(O_v)$ requires evaluating the APD of each local partition, which is computationally expensive. Therefore, \sys estimates APD through object sampling and reuses the sampled statistics of child nodes whenever possible, avoiding redundant computations and reducing the construction overhead.

\section{Cost-based Adaptive Search} \label{sec:online-search}

Given an MR-ANNS query $Q=(\qvec,\qattr,k)$, query processing should identify a set of local \hnsw{} graphs whose covered object subsets collectively contain all objects satisfying the range constraints.
Different node selections, however, lead to different search behaviors. The key challenge is therefore to determine an appropriate combination of nodes that balances the search cost on individual local graphs and the number of local graphs to search.

To better understand this trade-off, we first present two straightforward search strategies as illustrated in Figure~\ref{fig:rtreehnsw}, and then introduce our cost-based adaptive search algorithm.

\stitle{Single-Node Local Graph Search.}
Since each internal node covers all objects contained in its descendants, one straightforward strategy is to perform MR-ANNS on only a single local \hnsw{} graph. 
Specifically, \sys traverses the R-tree in a top-down fashion and selects the node with the smallest object subset that still covers all objects satisfying the range constraints.
It then performs MR-ANNS on the corresponding local \hnsw{} graph.
The advantage of this strategy is searching only one local \hnsw{} graph. However, the selected node usually contains many objects outside the query ranges, causing more irrelevant objects to be visited during graph exploration and thus increasing the search cost.

\stitle{All-Leaf-Nodes Local Graph Search.}
The other extreme is to perform MR-ANNS on the local \hnsw{} graphs of all leaf nodes whose \mbr{}s intersect the query ranges, and merge the retrieved candidates to produce the final result.
This strategy maximizes the pruning capability of the R-tree, since each leaf node contains only a small number of objects outside the query ranges. However, it requires searching many local \hnsw{} graphs, resulting in considerable graph traversal and coordination overhead.

\stitle{Our Approach.} 
To balance the trade-off, we propose a \emph{cost-based adaptive search} algorithm. Specifically, we first develop a cost model for MR-ANNS on a local \hnsw{} graph by considering both graph properties (maximum degree and graph height) and query-dependent factors (candidate-list length and range selectivity). 
We regard the user-provided parameter $ef_{\mathrm{search}}$ as the global search budget and seek a cost-effective execution plan by selecting an appropriate combination of local graphs and allocating the budget among them, while guaranteeing complete coverage of the query constraints.

To this end, we first develop a cost model to estimate the search cost of each candidate local \hnsw{} graph (Section~\ref{subsec:cost-model}), and then present a dynamic programming algorithm to jointly optimize local graph selection and search budget allocation (Section~\ref{subsec:node-select}).

%

\subsection{Local MR-ANNS Cost Model}\label{subsec:cost-model}
The search cost of MR-ANNS on a local \hnsw{} graph is dominated by high-dimensional vector distance computations, while the costs of other operations, such as range checking and merging local results, can be negligible. Therefore, we measure the search cost by the number of vector distance computations.

\stitle{Cost Formulation.}
Intuitively, the cost of performing MR-ANNS on a local \hnsw{} graph consists of two components: (1) traversing the upper layers to locate an entry point of the bottom layer, and (2) performing a best-first search on the bottom layer. The former mainly depends on the graph structure, whereas the latter is additionally affected by the local search budget $ef_v$ and the selectivity of the query ranges, denoted by $\operatorname{sel}_v(\qattr)$.
Formally, given a candidate node $v$ whose \mbr{} intersects the query ranges, we formulate upper-layer traversal cost and bottom-layer search cost as follows. 

\etitle{Upper-layer Traversal Cost.}
MR-ANNS first greedily descends from the top layer to the bottom layer to locate an entry point. Following the standard HNSW search analysis, each upper layer has an average degree of approximately $M_v/2$, and the graph contains $h_v$ layers in total. Therefore, the traversal cost is
estimated by $\left(\frac{M_v}{2}\right)^2 h_v$.

\etitle{Bottom-layer Search Cost.}
Starting from the entry point, MR-ANNS performs a best-first search on the bottom layer with local search budget $ef_v$. Since only objects satisfying the query ranges can be considered as candidates, approximately $ef_v/\operatorname{sel}_v(\qattr)$ objects need to be explored to obtain $ef_v$ valid candidates. Each explored object visits approximately $M_v$ neighboring vertices. Therefore, the bottom-layer search cost is estimated by
$\min\!\left\{
|O_v|,
\frac{M_vef_v}{\operatorname{sel}_v(\qattr)}
\right\}$.

Combining the above two components, we estimate the search cost of performing query $Q$ on node $v$ with budget $ef_v$ as
\begin{equation}
\label{eq:querying-local-cost}
\widehat{C}_v(Q,ef_v)
=
\left(\frac{M_v}{2}\right)^2h_v
+
\min\!\left\{
|O_v|,
\frac{M_vef_v}{\operatorname{sel}_v(\qattr)}
\right\}.
\end{equation}

\stitle{Range Selectivity Estimation.}
The remaining problem is to estimate the local range selectivity $\operatorname{sel}_v(\qattr)$ efficiently. Computing the exact selectivity requires scanning the local object subset $O_v$, which is prohibitively expensive during query processing.

Instead of estimating the selectivity directly from the \mbr{} of every node, \sys assumes local uniformity only at leaf nodes, where each partition is sufficiently compact. The selectivity of internal nodes is then recursively aggregated from their child nodes, yielding more accurate estimates than directly computing overlap ratios on higher-level \mbr{}s.

Specifically, we present the estimation methods in leaf and intermediate nodes as follows. 

\etitle{Leaf Nodes.}
For a leaf node $z$, we assume objects are uniformly distributed within its \mbr{} $R_z$. The range selectivity is therefore estimated by the overlap ratio between the query ranges and the \mbr{}, computed as the product of the overlap ratios on all attributes:
\begin{equation}
\label{eq:querying-leaf-selectivity}
\widehat{\operatorname{sel}}_z(\qattr)
=\prod_{j=1}^{m}\frac{|\qattr_j\cap R_{z,j}|}{|R_{z,j}|},
\end{equation}
where $|I|$ denotes the length of an interval $I$.

\etitle{Internal Nodes.}
For an internal node $v$, the range selectivity is recursively computed as the object-cardinality weighted average of the estimated selectivities of its two child nodes:
\begin{equation}
\label{eq:querying-internal-selectivity}
\widehat{\operatorname{sel}}_v(\qattr)
=
\frac{|O_{lch_v}|}{|O_v|}
\widehat{\operatorname{sel}}_{lch_v}(\qattr)
+
\frac{|O_{rch_v}|}{|O_v|}
\widehat{\operatorname{sel}}_{rch_v}(\qattr).
\end{equation}

\stitle{Search Budget Allocation.}
The local MR-ANNS cost model requires the search budget $ef_v$ for each candidate node. Starting from the user-provided global search budget $ef_{\mathrm{search}}$ at the root node, \sys recursively allocates the budget to the predicate-intersecting child nodes in proportion to their estimated numbers of qualifying objects. 
Specifically, let $\widehat{N}_v(\qattr) = |O_v|\widehat{\operatorname{sel}}_v(\qattr)$ denote the estimated number of objects satisfying the query ranges in node $v$. Then, for each child node $u\in\{lch_v,rch_v\}$, the local search budget is computed by
\begin{equation}
\label{eq:querying-ef-allocation}
ef_u
=
\operatorname{Round}
\!\left(ef_v\cdot
\frac{\widehat{N}_u(\qattr)}
{\widehat{N}_v(\qattr)}
\right).
\end{equation}
This allocation strategy assigns a larger search budget to partitions expected to contain more qualifying objects. Since $\widehat{N}_v(\qattr)=\widehat{N}_{lch_v}(\qattr)+ \widehat{N}_{rch_v}(\qattr)$, the allocated budget is naturally bounded by its parent budget, i.e., $ef_u\le ef_v$.

\begin{algorithm}[t]
\caption{Cost-based Adaptive Search}
\label{alg:cost-based-adaptive-search}
\begin{algorithmic}[1]
\Require \sys $T_{v_1,V}$, query $Q=(\qvec,\qattr,k)$, global candidate-list length $ef_{\mathrm{search}}$
\Ensure Answer $\widehat{S}_k(Q)$


\Function{DP}{$v$}
    \State $lcost\gets0,\quad lnodes\gets\emptyset$
    \If{$\neg lch_v\mathrm{\ is\ NULL}$}
        \State $lcost,lnodes\gets \Call{DP}{lch_v}$
    \EndIf
    \State $rcost\gets0,\quad rnodes\gets\emptyset$
    \If{$\neg rch_v\mathrm{\ is\ NULL}$}
        \State $rcost,rnodes\gets \Call{DP}{rch_v}$
    \EndIf
    \If{$lcost+rcost<\widehat{C}_v(Q,ef_v)$}
        \State \Return $lcost+rcost,lnodes\cup rnodes$
    \Else
        \State \Return $\widehat{C}_v(Q,ef_v),\{v\}$
    \EndIf
\EndFunction
\State $T_{v_1,V(\qattr)}\gets \Call{\mbr{}pruning}{T_{v_1,V},\qattr}$
\State $\Call{SelEst}{v_1,\qattr}$ to work out $s_v(\qattr)$ for each $v\in V(\qattr)$
\State $\Call{EfTuning}{v_1,\qattr,ef_{\mathrm{search}}}$ to set $ef_v$ for each $v\in V(\qattr)$
\State $cost,nodes\gets\Call{DP}{v_1}$
\State $\widehat{S}_k(Q)\gets \mathrm{top-}k\mathrm{\ nearest\ of\ }\bigcup_{v\in nodes}\Call{LocalGraphSearch}{v,Q,ef_v}$
\State \Return $\widehat{S}_k(Q)$
\end{algorithmic}
\end{algorithm}

\subsection{Cost-based Adaptive Search Optimization}\label{subsec:node-select}
Given the estimated local MR-ANNS cost for each node, we next optimize the problem that determines the node combination that minimizes the overall search cost while guaranteeing coverage of the query ranges. We formulate this optimization problem, and solve the optimization using a dynamic programming algorithm.

\stitle{Optimization Problem.}
After \mbr{} pruning, each candidate node $v$ is associated with an estimated local search cost $\widehat{C}_v(Q,ef_v)$. The objective is to find a subset of candidate nodes that minimizes the total search cost, i.e., 
\[
\min_{V_s\subseteq V(\qattr)}
\sum_{v\in V_s}\widehat{C}_v(Q,ef_v),
\]
subject to the constraint that exactly one node is selected on every root-to-leaf path of the pruned R-tree.

The above constraint guarantees both coverage of the query ranges and non-overlapping searches. Selecting no node on a root-to-leaf path may miss objects satisfying the query ranges, whereas selecting multiple nodes on the same path results in redundant searches over the same object subset.

\stitle{Idea of Dynamic Programming Optimization.}
A straightforward solution is to enumerate all feasible node combinations and evaluate their total search costs. However, such an approach is computationally prohibitive because the number of feasible combinations grows exponentially with the size of the pruned R-tree.
Fortunately, by exploiting the hierarchical structure of the R-tree, the optimization problem exhibits an optimal substructure, enabling an efficient dynamic programming solution.

For each candidate node $v$, there are only two possible choices.
\begin{itemize}[leftmargin=*]
\item \textbf{Select $v$.} Execute MR-ANNS on the local graph of $v$ with cost $\widehat{C}_v(Q,ef_v)$. Since $O_v$ covers all objects in its subtree, none of its descendants should be selected.
\item \textbf{Optimize the child subtrees.} Skip node $v$ and recursively optimize the two child subtrees. Since $O_{lch_v}\cap O_{rch_v}=\emptyset$, the two subproblems are independent, and their optimal solutions can be combined directly.
\end{itemize}
Therefore, the optimal search cost rooted at node $v$ is computed as
\begin{equation}
\label{eq:local-graph-combination-dp}
OPT(v)
=
\min\!\left\{
\widehat{C}_v(Q,ef_v),\;
OPT(lch_v)+OPT(rch_v)
\right\},
\end{equation}
where the first term corresponds to selecting node $v$, and the second term corresponds to recursively optimizing its two child subtrees.

Therefore, the optimal node combination is obtained by evaluating $OPT(v_1)$ in a bottom-up traversal of the pruned R-tree.

\stitle{DP-based Query Processing Algorithm.}
\Cref{alg:cost-based-adaptive-search} outlines the query processing procedure. Given an \sys{} index $T_{v_1,V}$, an MR-ANNS query $Q=(\qvec,\qattr,k)$, and the user-provided global search budget $ef_{\mathrm{search}}$, the algorithm first traverses the index from the root and prunes nodes whose \mbr{}s are disjoint from the query ranges, producing the pruned tree $T_{v_1,V(\qattr)}$.

For each node in the pruned tree, the algorithm estimates its local range selectivity. As discussed previously, the selectivity of a leaf node is estimated from the overlap between its \mbr{} and the query ranges according to \Cref{eq:querying-leaf-selectivity}, while the selectivity of an internal node is recursively aggregated from its child nodes according to \Cref{eq:querying-internal-selectivity}. Starting with $ef_{v_1}=ef_{\mathrm{search}}$, the algorithm further allocates a local search budget $ef_v$ to each candidate node based on its estimated number of objects satisfying the query ranges, following \Cref{eq:querying-ef-allocation}. With the estimated selectivities and allocated search budgets, the local MR-ANNS cost of each candidate node is computed using \Cref{eq:querying-local-cost}. The algorithm then evaluates the dynamic programming formulation in \Cref{eq:local-graph-combination-dp} in a bottom-up manner to determine the node combination with the minimum estimated search cost.

Finally, MR-ANNS is executed on the local \hnsw{} graphs of the selected nodes using their respective search budgets $ef_v$. During each local search, only objects satisfying the query ranges are retained as result candidates. The candidates returned from all selected graphs are then merged and ranked by their distances to $\qvec$, producing the final approximate top-$k$ result $\widehat{S}_k(Q)$.

\stitle{Complexity Analysis.}
The proposed optimization introduces only linear overhead with respect to the size of the pruned R-tree. Specifically, the dynamic programming algorithm evaluates each node exactly once, resulting in a time complexity of $\mathrm{O}(|V(\qattr)|)$. Moreover, the preprocessing steps, including range selectivity estimation and local search budget allocation, each traverse the pruned tree once and therefore also run in $\mathrm{O}(|V(\qattr)|)$ time. In practice, \mbr{} pruning significantly reduces the search space, yielding $|V(\qattr)|\ll|V|\ll|O|$. 
As a result, the optimization overhead is negligible compared with the subsequent MR-ANNS execution.

\section{Incremental Index Update} \label{sec:index-update}


\begin{figure}
    \centering
    \includegraphics[width=\columnwidth]{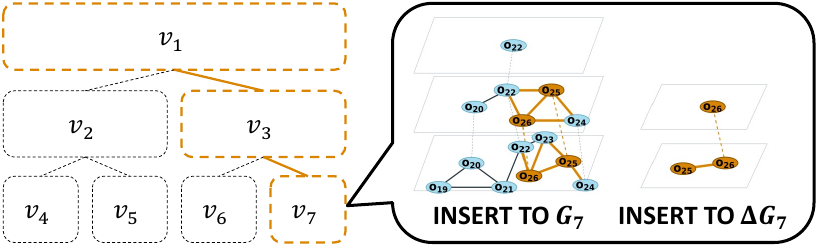}
    \caption{Comparison of Main vs. Delta Graph Insertions.}
    \label{fig:indexupdates-1}
\end{figure}

Maintaining \sys{} under object insertions and deletion requires updating both the R-tree for attributes and local \hnsw{} graphs for vectors.
While the R-tree can be updated efficiently, incrementally maintaining the local \hnsw{} graphs is more expensive because inserting new objects triggers extensive edge modifications.
Specifically, although HNSW supports incremental insertion, a straightforward solution inserts each new object into every affected local graph. As illustrated in \Cref{fig:indexupdates-1}, inserting only two new objects into $\vhnsw_7$ already causes more edge modifications between the new and existing objects, resulting in high update overhead.

To address this, \sys adopts a \emph{delta-based incremental update} mechanism. Instead of directly modifying the original local graphs, each node maintains an additional sparse delta graph (e.g., $\Delta\vhnsw_7$ in \Cref{fig:indexupdates-1}) to accommodate newly inserted objects. Consequently, updates only affect the lightweight delta graphs, substantially reducing graph maintenance cost. During query processing, MR-ANNS is executed on both the original and delta graphs, and their results are merged to produce the final top-$k$ answer. The remainder of this section presents the index update with delta graphs and the corresponding query processing algorithm.

%
%

\subsection{Delta-based Incremental Update}
Incremental updates in \sys{} consist of two components: updating the R-tree for attributes and maintaining the local \hnsw{} graphs for vectors.
While attribute updates only modify the \mbr{}s along the insertion path, vector updates are more challenging because incrementally inserting objects into the local \hnsw{} graphs triggers extensive edge modifications. 

To address this issue, \sys{} adopts a delta-based incremental update mechanism, which maintains sparse delta graphs for newly inserted objects while keeping the original graphs unchanged. Besides, object deletions are handled lazily using a bitmap without modifying the index structure. Deleted objects are filtered out during subsequent MR-ANNS according to their bitmap flags.
For clarity, we refer to the original local \hnsw{} graph over object set $O_v$ as the \emph{main graph} $\vhnsw_v$, and the auxiliary graph over newly inserted objects $\Delta O_v$ as the \emph{delta graph} $\Delta\vhnsw_v$. The \mbr{} of each node always bounds the attributes of $O_v\cup\Delta O_v$.

\stitle{Attribute Update.}
Given a newly inserted object $o_i=(\ovec_i,\oattr_i)$, \sys{} first locates its root-to-leaf insertion
path according to its attribute values. Along the insertion path, $o_i$ is inserted into the corresponding delta object set $\Delta O_v$. If $\oattr_i$ falls outside the current \mbr{} $R_v$ of a visited node, $R_v$ is expanded to enclose the new object.

Unlike conventional R-trees, \sys{} does not perform node splitting or merging after insertions. Since they requires reconstructing all local \hnsw{} graphs in the affected subtree, avoiding structural modifications significantly reduces the update overhead.

\stitle{Vector Update.}
Instead of directly inserting new objects into main graph $\vhnsw_v$, \sys{} incrementally maintains a sparse delta graph $\Delta\vhnsw_v$ for each node. Since newly inserted objects typically
constitute only a small fraction of the indexed objects, all insertions are absorbed by the lightweight delta graphs while the main graphs remain unchanged.
After locating the insertion path, the new object is inserted into the delta graph of every visited node. To reduce the maintenance cost, the delta graph is constructed with a smaller maximum degree: $M_v^\Delta = \operatorname{Round}\!\left(\frac{M_v}{\alpha}\right)$.

Thus, inserting an object into $\Delta\vhnsw_v$ requires substantially fewer neighbor searches and edge
modifications than inserting it into the main graph, as illustrated in \Cref{fig:indexupdates-1}.

\stitle{Update Complexity.}
Given the local HNSW graph construction parameter $ef_{\mathrm{construction}}=\omega_{c,v}$, inserting an object into the main graph $\vhnsw_v$ incurs a cost of approximately $\left(\frac{M_v}{2}\right)^2h_v+M_v\omega_{c,v}$,
whereas inserting it into the delta graph incurs $\left(\frac{M_v^\Delta}{2}\right)^2h_v^\Delta
+
M_v^\Delta\omega_{c,v}$.

Since $M_v^\Delta=M_v/\alpha$ and $\omega_{c,v}\propto M_v$ in practice, the update cost on the delta graph is approximately $1/\alpha^2$ of that on the main graph. Therefore, the proposed delta-based incremental update reduces the graph maintenance overhead by roughly a factor of $\alpha^2$.

\subsection{Search over Main and Delta Graphs}

After the delta-based incremental update, the attribute index remains unchanged, whereas the vector index is split into two graph structures. 
Specifically, the \mbr{} $R_v$ of each node still bounds the attributes of $O_v\cup\Delta O_v$, allowing the standard \mbr{} pruning procedure to identify all candidate nodes without missing newly inserted objects.
The vectors, however, are indexed separately by the main graph $\vhnsw_v$ over $O_v$ and the delta graph $\Delta\vhnsw_v$ over $\Delta O_v$. Therefore, querying only the main graph would miss newly inserted objects, making it necessary to jointly search both \hnsw{} graphs.

For each selected node $v$, the main graph is searched using its local candidate-list length $ef_v$. To compensate for the separate navigation on the delta graph, \sys{} allocates it an additional candidate-list length proportional to its relative size:
\begin{equation}
\label{eq:deltagraphef}
ef_v^\Delta
=
\operatorname{Round}
\!\left(
ef_v
\frac{|\Delta O_v|}{|O_v|}
\right).
\end{equation}
This proportional allocation assigns more search effort to larger delta graphs while introducing only limited overhead when few objects have been inserted.

\sys{} performs MR-ANNS on the main graph $\vhnsw_v$ with budget $ef_v$ and on the delta graph $\Delta\vhnsw_v$ with budget $ef_v^\Delta$. During both searches, an object is retained as a candidate only if it satisfies all range constraints and is not marked in the deletion bitmap. The candidates returned from the two graphs at all selected nodes are then merged and ranked by their Euclidean distances to $\qvec$, producing the final approximate top-$k$ result.

Compared with searching only the main graph, jointly searching the delta graph introduces the following additional search cost:
\begin{equation}
\label{eq:querying-deltagraph-cost}
\widehat{C}_v^\Delta(Q,ef_v^\Delta)
=
\left(\frac{M_v^\Delta}{2}\right)^2h_v^\Delta
+
\min\!\left\{
|\Delta O_v|,
\frac{M_v^\Delta ef_v^\Delta}
{\widehat{\operatorname{sel}}_v(\qattr)}
\right\}.
\end{equation}

\stitle{Adaptive Delta Merging.}
Although the sparse delta graph significantly reduces update overhead, it also introduces additional query cost. Moreover, as insertions accumulate, the delta graph grows larger, making joint search
progressively more expensive. To balance update and query efficiency, \sys{} adopts an adaptive delta merging strategy based on cost comparison.

Specifically, for each node $v$, \sys{} accumulates the additional search cost incurred on the delta graph over all queries, denoted by $\Sigma\widehat{C}_v^\Delta$, and compares it with the one-time cost of merging the delta graph into the main graph:
\begin{equation}
\label{eq:onetime-merge-cost}
\widehat{C}_v^{\mathrm{merge}}
=
|\Delta O_v|
\left[
\left(\frac{M_v}{2}\right)^2h_v
+
M_v\omega_{c,v}
\right].
\end{equation}

A merge is triggered once $\Sigma\widehat{C}_v^\Delta > \widehat{C}_v^{\mathrm{merge}}$.
After merging, all unmarked objects in $\Delta O_v$ are incrementally inserted into the main graph $\vhnsw_v$, after which $\Delta\vhnsw_v$ and $\Delta O_v$ are cleared. Subsequent insertions are accumulated in a new delta graph, forming a repeated cycle of delta-based updates, joint search, and adaptive merging.

\section{Experiments}
\label{sec:experiments}

We conduct extensive experiments to evaluate \sys from three aspects: query performance, incremental update, and index construction. The remainder of this section is organized accordingly. We first evaluate the query performance of \sys, followed by the evaluation of its incremental update mechanism, and finally its index construction efficiency.



\subsection{Experiment Setup}
\label{sec:exp-setup}

\noindent\textbf{Datasets.}
We evaluate all methods on three real-world public datasets: DBLP~\cite{dblp}, MSMarco~\cite{msmarco}, and LAION~\cite{laion}, covering diverse application domains, vector dimensions, and multi-attribute range query characteristics. Their statistics are summarized in Table~\ref{tab:datasets}.

\begin{itemize}[leftmargin=*]
    \item \textbf{DBLP} is a scholarly publication dataset with 768-dimensional document embeddings. We use publication metadata, including publication year and statistics of citations, references, and authors, as range query attributes.

    \item \textbf{MSMarco} is a large-scale text retrieval dataset with 384-dimensional document embeddings. We use document statistics, including the numbers of words, characters, sentences, unique words, and TF-IDF scores, as range query attributes.

    \item \textbf{LAION} is a large-scale image-text dataset with 512-dimensional image-text embeddings. We use image metadata, including image width, image height, and image-text similarity scores, as range query attributes.
\end{itemize}

For all datasets, we adopt the vector representations and associated metadata constructed in prior work~\cite{yu2026efficient}. To support multi-attribute numerical range queries, we transform metadata into numerical attributes whenever necessary. Specifically, for metadata that are not directly numerical, we derive numerical statistics from their associated information (e.g., the numbers of citations, references, authors, words, and sentences). This preprocessing yields a unified benchmark with numerical range attributes while preserving the original semantics of the metadata.

\stitle{Queries.}
To comprehensively evaluate MR-ANNS under different filtering conditions, we generate query workloads covering a broad spectrum of selectivities. This design is also motivated by real-world multidimensional range workloads, whose query selectivities often span several orders of magnitude~\cite{sprenger2018multidimensional,liu2025good}.

For each query, we uniformly sample the number of range constraints $p$ from $[1,m]$ and randomly select $p$ attributes without replacement. 
We then sample a desired query selectivity $s$ log-uniformly from $[10^{-4},1]$, so that query selectivities span multiple orders of magnitude and each order of magnitude is approximately equally represented. Assuming attribute independence only for workload generation, each selected attribute is assigned an expected selectivity of $s^{1/p}$.
Next, we randomly sample an object, use its vector as the query vector, and construct a range constraint around the corresponding value of each selected attribute. The range boundaries are randomly perturbed to ensure query diversity. Queries returning fewer than $k$ objects are discarded and regenerated. The exact filtered top-$k$ results are computed as the ground truth. Unless otherwise specified, we generate 1,000 queries per dataset, set $k=10$, and use Euclidean distance.


\stitle{Evaluation Metrics.}
Following prior ANNS work, we use the QPS--Recall curve as the primary evaluation metric, where Recall is computed against the exact filtered top-$k$ results and QPS measures query throughput. Curves closer to the upper-right corner indicate better search performance.

\begin{table}[t!]
    \centering
    \renewcommand{\arraystretch}{1.0}
    \setlength{\tabcolsep}{12pt}
    \caption{Statistics of the datasets. Here, $n$, $d$, and $m$ denote the number of objects, the vector dimensionality, and the number of numerical attributes, respectively.}
    \label{tab:datasets}
    \begin{tabular}{lrrrc}
        \toprule
        Dataset & $n$ & $d$ & $m$ & Data Type \\
        \midrule
        DBLP    & 6,275,270 & 768 & 4 & Text  \\
        MSMarco & 8,000,000 & 384 & 5 & Text  \\
        LAION   & 9,636,707 & 512 & 3 & Image \\
        \bottomrule
    \end{tabular}
    \vspace{-1em}
\end{table}

\stitle{Methods and Settings.}
We compare \sys with three representative MR-ANNS methods, \textbf{DIGRA}~\cite{jiang2025digra}, \textbf{KHI}~\cite{yu2026efficient}, and \textbf{ACORN-$\gamma$}~\cite{patel2024acorn}, together with two baselines, \textbf{Pre-filtering} and \textbf{Post-filtering}.
Pre-filtering first evaluates the range constraints and then performs exact brute-force nearest neighbor search over the qualified objects, whereas Post-filtering first performs ANN search and subsequently filters out objects that do not satisfy the range constraints.
All graph-based methods are built upon the NSW/HNSW framework. We use $M$ to denote the maximum graph degree and $\omega_c$ the graph construction beam width. Unless otherwise specified, all baselines use the parameter settings recommended in their original papers.



\begin{itemize}[leftmargin=*]
    \item \textbf{\sys} is our proposed method. We set the leaf capacity to $0.1\%|O|$, the adaptive degree range to $M_L=24,M_H=32$, and the construction beam width of each local graph $\vhnsw_v$ to $\omega_{c,v}=8M_v$.
    %
    \item \textbf{DIGRA} organizes objects by a single numerical attribute using a dynamic multi-way tree and maintains node-level NSW graphs. Since DIGRA supports only single-attribute filtering, we index one selected attribute and treat the remaining range constraints as post-filters. We set $M=32$ and $\omega_c=400$.
    %
    \item \textbf{KHI} partitions the multi-dimensional attribute space using a KD-tree and maintains local HNSW graphs for individual partitions. We set $M=32$ and $\omega_c=128$.
    %
    \item \textbf{ACORN-$\gamma$} maintains a range-agnostic HNSW graph and performs search over range-induced subgraphs. We set $M=32$, $M_{\beta}=64$, $\gamma=100$, and $\omega_c=M\gamma$. The original parameter rule recommends $\gamma=1/s_{\min}$, which gives $\gamma=10^4$ for our minimum query selectivity and leads to prohibitive construction cost. We therefore use $\gamma=100$, corresponding to a selectivity threshold of approximately $1\%$, as a practical trade-off between construction cost and query performance.
    %
    \item \textbf{Pre-filtering and Post-filtering}. Pre-filtering builds an R-tree over the numerical attributes and performs exact search over the qualified objects. Post-filtering builds a global HNSW graph and filters the returned objects using the range constraints. We set $M=32$ and $\omega_c=64$ for Post-filtering.
\end{itemize}


Note that, for DIGRA and Post-filtering, we additionally increase the intermediate candidate set size up to $10{,}000$ (i.e., $1{,}000k$ for $k=10$) to fully exploit their search capability and achieve recall comparable to the other methods.

\stitle{Experiment Settings.}
All experiments were conducted on a server equipped with Intel(R) Xeon(R) Gold 6230 CPUs at 2.10\,GHz and 1\,TiB of RAM, running Ubuntu 20.04.6 LTS. All methods were implemented in C++ and compiled using GCC 14.3.0. To ensure a fair comparison, all methods were evaluated under the same hardware environment, datasets, query workloads, and execution settings. We ran each experiment three times and reported the average result.

\begin{figure*}[t]
	\centering
	\includegraphics[width=1\textwidth]{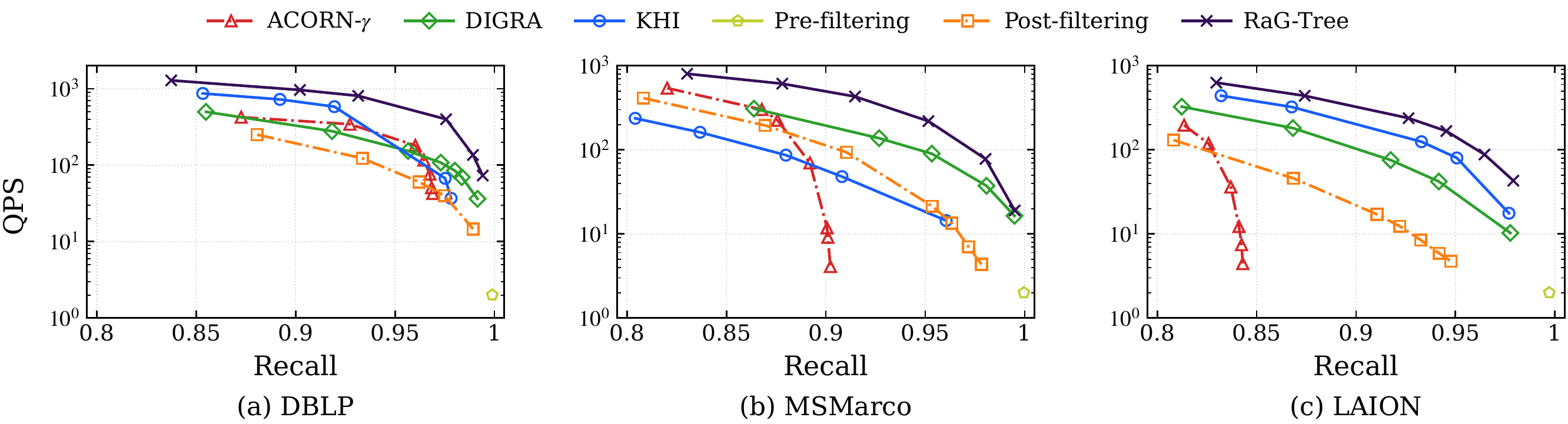}
	\caption{Overall QPS vs. recall performance comparison of \sys
    and the baselines on three datasets.} 
	\label{fig:overall-performance}
\end{figure*}

\subsection{Search Performance}
\label{sec:search-performance}
Figure~\ref{fig:overall-performance} shows the QPS-Recall curves of all methods. Overall, \sys achieves the best QPS--Recall trade-off across all three datasets. For example, at Recall around $0.95$, \sys
achieves approximately $2.8\times$, $2.4\times$, and $1.9\times$ higher QPS than the strongest competing method on DBLP, MSMarco, and LAION, respectively. 
Next, we analyze these observations in more detail.

The comparison with ACORN-$\gamma$ demonstrates that improving MR-ANNS solely through graph exploration is insufficient. Under the practical $\gamma=100$ setting, ACORN-$\gamma$ is less competitive on MSMarco and LAION, where multi-attribute range queries typically exhibit low selectivities, requiring pruning irrelevant objects before graph exploration. Although increasing graph connectivity can improve graph navigation under filtering, it also incurs higher graph construction and traversal costs. 

DIGRA further shows the limitation of partitioning the attribute space using only a single attribute. It organizes objects according to one selected numerical attribute, while the remaining range constraints are evaluated through post-filtering. As the number of query attributes increases, an increasingly larger fraction of out-of-range objects remain in the searched partitions, leading to unnecessary graph exploration and vector distance computations. These results indicate that effective MR-ANNS requires jointly considering multiple attribute dimensions during range pruning.

KHI extends attribute partitioning to multiple dimensions by combining a KD-tree with node-level HNSW graphs. However, it still does not consistently outperform the other baselines on DBLP and MSMarco. This is because the KD-tree partitions the attribute space along one dimension at each split and therefore cannot effectively capture the correlations among multiple attributes. Consequently, multi-attribute range queries may still intersect multiple partially relevant partitions, triggering redundant local graph searches. Moreover, KHI selects local graphs primarily according to range coverage under a predefined node limit, without considering their search costs. These observations suggest that multi-dimensional partitioning alone is insufficient, and the query execution strategy over local graphs is important.

The superior performance of \sys stems from integrating the above components into a unified design. First, the R-tree organizes objects in the multi-dimensional attribute space, allowing each MBR to jointly capture all attribute dimensions and better preserve the underlying attribute correlations, thereby improving range pruning effectiveness.
Second, the correlation-aware local HNSW graphs exploit the attribute-vector correlations across partitions to improve graph exploration efficiency. 
Finally, cost-based adaptive search adaptively determines the local graph combination and allocates search budgets according to the estimated search costs, avoiding unnecessary graph exploration while maintaining high recall. Overall, the above designs enable \sys to consistently achieve the best QPS-Recall trade-off across all three datasets.

\begin{table}[t]
\centering
\caption{Effect of Cost-based Adaptive Search (CAS). 
}
\label{tab:budget-ablation}
\small
\setlength{\tabcolsep}{6pt}
\begin{tabular}{lrrrrr}
\toprule
& \multicolumn{2}{c}{\sys w/o CAS}
& \multicolumn{2}{c}{\sys} & \\
\cmidrule(lr){2-3}
\cmidrule(lr){4-5}
Dataset & Recall@10 & QPS & Recall@10 & QPS & QPS Imp. \\
\midrule
\multirow{3}{*}{DBLP}
& 0.8216 & 1,067 & 0.8374 & 1,285 & \textbf{20\%} \\
& 0.9236 &   462 & 0.9315 & 807  & \textbf{75\%} \\
& 0.9891 &    90 & 0.9892 & 136   & \textbf{51\%} \\
\midrule
\multirow{3}{*}{MSMarco}
& 0.8255 & 658 & 0.8302 & 802 & \textbf{22\%} \\
& 0.9516 & 109 & 0.9517 & 219 & \textbf{101\%} \\
& 0.9841 &  33 & 0.9858 & 53  & \textbf{61\%} \\
\midrule
\multirow{3}{*}{LAION}
& 0.8261 & 333 & 0.8297 & 629 & \textbf{89\%} \\
& 0.9443 &  44 & 0.9454 & 167 & \textbf{280\%} \\
& 0.9768 &  14 & 0.9791 & 43  & \textbf{207\%} \\
\bottomrule
\end{tabular}
\end{table}

\subsection{Impact of Cost-based Search Optimization}
\label{sec:budget-ablation}
To evaluate the impact of Cost-based Adaptive Search (CAS), we compare the complete \sys with \sys w/o CAS. Both variants use the same index and construction parameters, differing only in the search strategy. \sys w/o CAS always searches a single R-tree node whose MBR covers the query range and assigns the entire global search budget to its local HNSW graph. In contrast, \sys employs Cost-based Adaptive Search to select a cost-effective combination of R-tree nodes and distribute the global search budget among their local HNSW graphs.

Table~\ref{tab:budget-ablation} reports the results. \sys improves both search efficiency and search quality across all nine evaluated configurations. Specifically, it improves QPS by $20\%$--$280\%$, with an average speedup of $100.7\%$. Meanwhile, it also achieves slightly higher Recall in every setting, indicating that the throughput improvement is obtained without sacrificing search accuracy. In particular, the benefit becomes more significant on larger datasets. For example, on LAION, \sys improves QPS by $1.89\times$--$3.80\times$, demonstrating that cost-based search optimization becomes increasingly important as the search space grows.

The improvement comes from balancing the trade-off between range pruning and graph exploration. Searching only a single covering node reduces local graph invocations, but its local HNSW graph typically contains many objects outside the query range, resulting in unnecessary graph exploration. Conversely, searching more local graphs improves pruning effectiveness, but also incurs additional graph invocations. CAS explicitly models this trade-off using the proposed cost model, adaptively selecting the node combination with the minimum estimated search cost and allocating the search budget according to the estimated numbers of potentially in-range objects. Consequently, \sys achieves more effective range pruning while avoiding excessive graph exploration. 

\begin{figure*}[t]
    \centering
    \includegraphics[width=\textwidth]{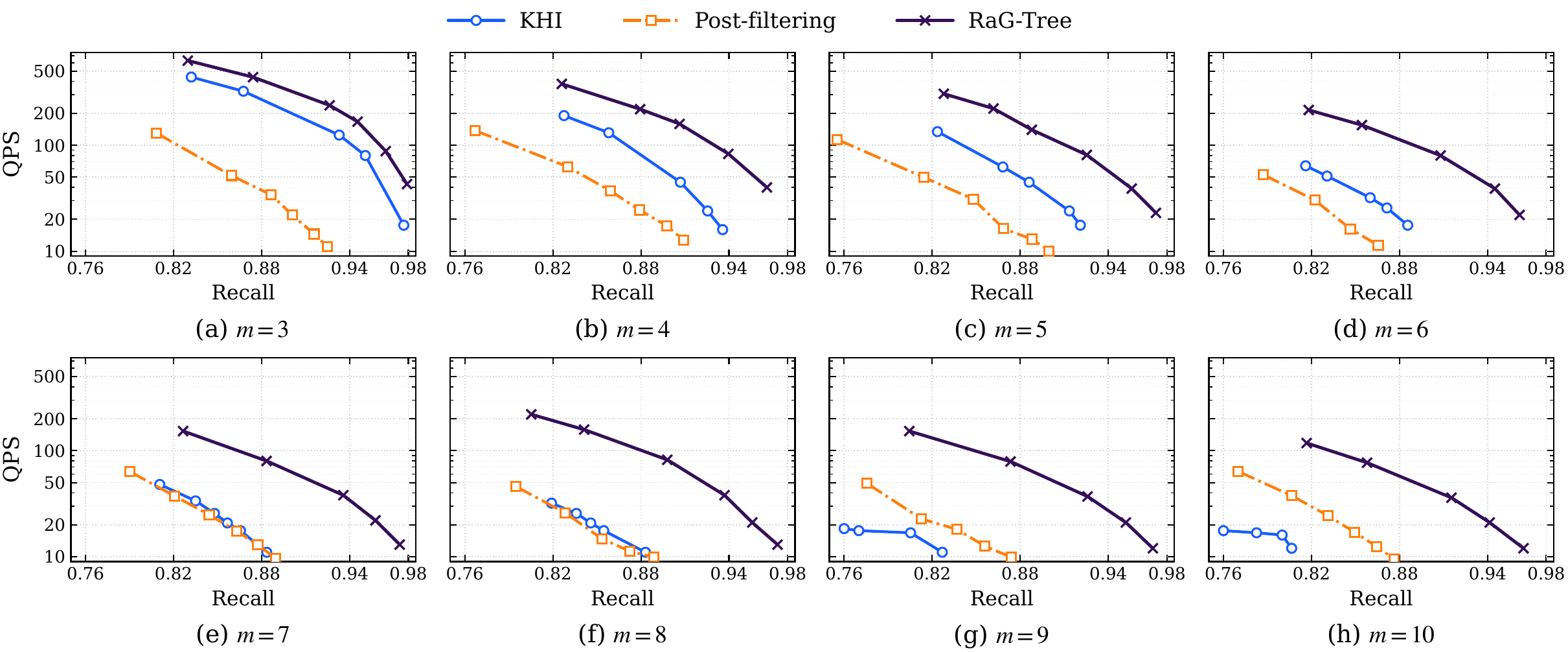}
    \caption{QPS vs. recall trade-offs of \sys and representative baselines with varying numbers of attributes on LAION.}
    \label{fig:khi-ragtree-varying-attributes}
\end{figure*}

\subsection{Impact of Attribute Dimensionality}
\label{sec:attribute-scalability}
We evaluate the scalability of \sys with respect to the number of attributes on LAION by increasing the attribute dimensionality from three to ten and comparing \sys with KHI and Post-filtering. Each additional attribute is generated by resampling an original attribute to preserve its marginal distribution while remaining approximately independent. We keep the vector embeddings and target selectivities fixed, and recompute the query constraints and exact ground truth for each setting.

Figure~\ref{fig:khi-ragtree-varying-attributes} reports the QPS-Recall curves under different numbers of attributes. All methods gradually degrade as attribute dimensionality increases. Nevertheless, \sys consistently achieves the best QPS-Recall trade-off. In contrast, Post-filtering mainly suffers a throughput reduction, while KHI has degradation in both throughput and recall. Specifically, at comparable recall levels, \sys achieves approximately $6.7\times$--$22.0\times$ higher QPS than Post-filtering and $1.9\times$--$11.0\times$ higher QPS than KHI.

The different performances reflect the distinct ways in which the methods handle increasingly selective multi-attribute range queries.
Post-filtering performs graph exploration without exploiting attribute constraints and therefore cannot reduce the search space as additional attributes are introduced. Consequently, more out-of-range objects are explored, resulting primarily in lower QPS.
KHI incorporates multi-attribute partitioning, but its KD-tree performs axis-aligned partitioning along one attribute at each split. As the number of attributes increases, a query tends to intersect more
partially relevant partitions, and KHI has to search coarser partitions that contain many out-of-range objects or reduce graph exploration, leading to losses in throughput and recall.

In contrast, \sys remains robust because its design scales naturally with increasing attribute dimensionality. The R-tree jointly organizes objects in the multi-dimensional attribute space, allowing each MBR to capture all attribute dimensions and preserve the underlying attribute correlations. This enables effective pruning even for high-dimensional range queries. Moreover, cost-based adaptive search adaptively determines the local graph combination and allocates search budgets according to the estimated search costs, maintaining an effective balance between range pruning and graph exploration.

\begin{figure}[t]
	\centering
	\includegraphics[width=\columnwidth]{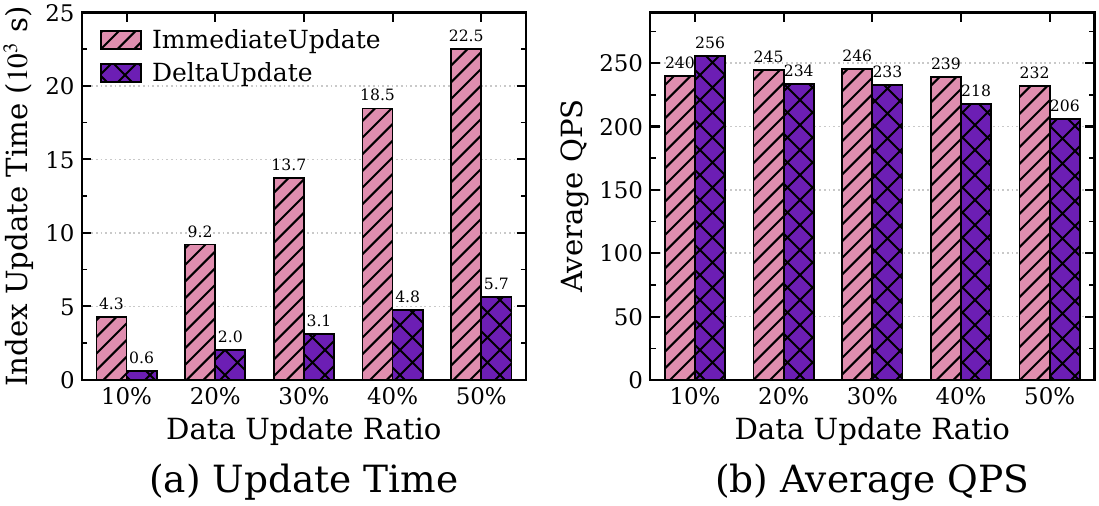}
    \vspace{-2em}
	\caption{Update time and average QPS of ImmediateUpdate and DeltaUpdate under different update ratios on LAION.}
    \label{fig:update}
    \vspace{-1em}
\end{figure}

\subsection{Evaluation of Index Updating}
\label{sec:update-performance}
We evaluate index updating on LAION with update ratios ranging from $10\%$ to $50\%$. Each update workload contains equal numbers of insertions and deletions in a randomly shuffled order. Inserted objects are generated by perturbing the vectors of existing objects to avoid duplicates. After each update workload, we recompute the ground-truth results to ensure accurate recall evaluation. 

We compare two update strategies. 
\emph{ImmediateUpdate} inserts every new object directly into the main \hnsw{} graph at each affected node.
\emph{DeltaUpdate} ($\alpha=2$ by default) instead inserts new objects into the corresponding delta \hnsw{} graphs and incrementally merges them into the main graphs through adaptive delta merging. 
Both strategies use the same deletion bitmap and differ only in how insertions are handled.

Figure~\ref{fig:update}(a) compares the update efficiency of the two strategies. DeltaUpdate consistently outperforms ImmediateUpdate across all evaluated update ratios, achieving a $3.87\times$--$6.81\times$ speedup with an average of $4.72\times$. 
ImmediateUpdate performs graph traversal, neighbor selection, and edge maintenance directly on the main
\hnsw{} graph for every inserted object. 
In contrast, DeltaUpdate first accumulates new objects in lightweight delta graphs and postpones the more expensive maintenance of the main graphs until adaptive delta merging is triggered. 
By amortizing the graph maintenance cost over multiple insertions, DeltaUpdate substantially reduces the update overhead. The improvement remains significant even under heavy update workloads. For example, at a $50\%$ update ratio, DeltaUpdate still reduces the update time by $74.8\%$.

Figure~\ref{fig:update}(b) compares the query performance after updates.
The average QPS is measured over five recall-aligned operating points, whose maximum Recall difference is only $0.004$. DeltaUpdate retains $88.7\%$--$95.4\%$ of the throughput of ImmediateUpdate for update ratios from $20\%$ to $50\%$, and even slightly outperforms it at the $10\%$ update ratio. The additional query cost comes from searching both the main and delta \hnsw{} graphs before merging their results. As more objects accumulate in the delta graphs, this overhead gradually increases. However, adaptive delta merging effectively bounds the overhead by merging a delta graph once its accumulated search cost exceeds the estimated one-time merging cost.

\begin{table}[t]
\centering
\caption{Index construction time (s) and index size (GB).}
\label{tab:indexing-overhead}
\small
\setlength{\tabcolsep}{6pt}
\resizebox{\columnwidth}{!}{
\begin{tabular}{lrrrr}
\toprule
\multicolumn{5}{c}{\textit{Index Construction Time (s)}} \\
\midrule
Dataset & ACORN-$\gamma$ & DIGRA & KHI & \sys \\
\midrule
DBLP & 3,246,360 & 248,580 & \underline{61,247} & \textbf{59,244} \\
MSMarco & 2,308,722 & 319,375 & \underline{103,803} & \textbf{66,110} \\
LAION & 2,569,788 & 298,624 & \underline{96,532} & \textbf{84,976} \\
\midrule\midrule
\multicolumn{5}{c}{\textit{Index Size (GB)}} \\
\midrule
Dataset & ACORN-$\gamma$ & DIGRA & KHI & \sys \\
\midrule
DBLP & 23.1 & 25.0 & \textbf{9.0} & \underline{9.2} \\
MSMarco & \underline{18.0} & 23.8 & 26.1 & \textbf{12.4} \\
LAION & 26.2 & 30.0 & \underline{21.3} & \textbf{15.6} \\
\bottomrule
\end{tabular}
}
\end{table}

\subsection{Evaluation on Index Construction}
\label{sec:construct-cost}
Table~\ref{tab:indexing-overhead} reports the index construction time and index size of all specialized filtered ANN indexes. Pre-filtering and Post-filtering are excluded because they do not construct MR-ANNS indexes.

\sys achieves the shortest index construction time on all three datasets. Compared with the fastest competitor (KHI), it reduces construction time by $3.3\%$, $36.3\%$, and $12.0\%$ on DBLP, MSMarco,
and LAION, respectively, corresponding to an average speedup of $1.25\times$. Although \sys builds a local \hnsw graph at every R-tree node, its partition-aware graph construction assigns smaller maximum
degrees to partitions whose local vectors are easier to navigate, thereby reducing the costs of neighbor selection and edge maintenance.
Moreover, the required attribute-vector correlations are estimated from sampled objects rather than exhaustive pairwise vector distances, introducing little additional overhead.

\sys also achieves the smallest index size on MSMarco and LAION. On DBLP, its $9.2$\,GB index is within $2.2\%$ of KHI, which achieves the smallest index size. 
The lightweight index mainly comes from the R-tree, which jointly partitions the multi-dimensional attribute space and therefore requires fewer local \hnsw graphs than alternative partition structures. In addition, partition-aware graph construction avoids assigning unnecessarily large maximum degrees to all local graphs, further reducing the number of stored graph edges.

\section{Related Work} \label{sec:rw}
\stitle{Approximate Nearest Neighbor Search} 
Approximate nearest neighbor search (ANNS) methods~\cite{li2019approximate, pan2024survey} can be broadly classified into hashing-based~\cite{andoni2008near, andoni2015practical, datar2004locality, wei2024det}, partition-based~\cite{andre2015cache, gao2024rabitq, ge2013optimized, jegou2010product, muja2014scalable}, and graph-based~\cite{fu2022high, fu2019fast, malkov2014approximate, malkov2018efficient, peng2023efficient, jayaram2019diskann} approaches. 
Hashing-based methods provide theoretical approximation guarantees but typically require substantial space to achieve high accuracy. Partition-based methods improve search efficiency by restricting queries to a small number of vector partitions, often combined with vector quantization for compact storage. Graph-based methods, such as HNSW~\cite{malkov2018efficient}, NSG~\cite{fu2019fast} and DiskANN~\cite{jayaram2019diskann}, organize vectors as proximity graphs and perform greedy graph
exploration, achieving an effective trade-off between search accuracy and efficiency.
These indexes target unconstrained ANNS and cannot exploit effectively attribute constraints for MR-ANNS.

\stitle{Attribute-Filtered Approximate Nearest Neighbor Search.}
Existing methods for attribute-filtered ANNS integrate attribute constraints into either query processing or index construction, and can generally be divided into filtering-based approaches and specialized indexes.
Filtering-based approaches combine ANN indexes with pre- or post-filtering and are widely adopted in vector databases such as Pgvector~\cite{PGVector}, Milvus~\cite{wang2021milvus}, and Qdrant~\cite{Qdrant}. They support flexible filters but handle attribute filtering and graph search independently.
%
Specialized indexes incorporate attribute information into the index structure and can be divided into predicate-agnostic and predicate-specific approaches. Predicate-agnostic indexes, such as ACORN~\cite{patel2024acorn}, Navix~\cite{sehgal2025navix}, and Rwalks~\cite{ait2025rwalks}, support diverse filters by augmenting proximity graphs or adapting graph traversal. Predicate-specific indexes target particular filter types: equality-filtered ANNS indexes~\cite{wang2023efficient, gollapudi2023filtered} focus on categorical attributes, while range-filtered ANNS indexes focus on numerical range constraints. Most existing range-filtered methods, including SeRF~\cite{zuo2024serf}, WST~\cite{engels2024approximate}, iRangeGraph~\cite{xu2024irangegraph}, WoW~\cite{wang2025wow}, and DIGRA~\cite{jiang2025digra}, support only single-attribute range filtering. KHI~\cite{yu2026efficient} extends this line of work to MR-ANNS by recursively partitioning with node-local HNSW graphs, still inefficient. 
In contrast, \sys jointly organizes multiple numerical attributes using an R-tree, exploits attribute and attribute-vector correlations, and optimizes graph exploration via cost-based adaptive search for more effective range pruning and graph exploration in MR-ANNS.

\section{Conclusion and Future Work}
In this paper, we have introduced \sys, a unified index for multi-attribute range approximate nearest neighbor search (MR-ANNS). \sys tightly couples an R-tree with partition-aware HNSW graphs, enabling efficient multi-dimensional range pruning and graph exploration within a unified framework. Specifically, the R-tree preserves attribute correlations through hierarchical multi-attribute partitions, while partition-aware HNSW graphs exploit local attribute-vector correlations to adapt graph sparsity to different partitions. Building upon this index structure, we have proposed cost-based adaptive search to optimize local graph selection and search budget allocation, together with a delta-based incremental update mechanism for efficient index maintenance under dynamic workloads. Extensive experiments on three real-world datasets have demonstrated that \sys consistently achieves the best QPS-Recall trade-off, while also providing efficient index construction, lightweight storage, and fast incremental updates. 

Several directions remain for future work. First, while this paper focuses on numerical attributes with range constraints, extending \sys to support heterogeneous filter types, such as categorical, textual, and spatial filters, is an important direction. Second, further integrating adaptive workload modeling into both index organization and query processing may enable \sys to optimize itself under evolving data distributions and query workloads.

\newpage
\bibliographystyle{ACM-Reference-Format}
\bibliography{reference}


%

\end{document}